\documentclass[aps,pra,showpacs, twocolumn
]{revtex4}
\usepackage{amssymb}
\usepackage{amsthm}
\usepackage{amsmath}
\usepackage[dvips]{epsfig}
\usepackage{graphicx}
\usepackage{color}
\usepackage{slashbox}

\begin{document}

\title{Oscillatory instabilities of gap solitons in a  repulsive Bose-Einstein condensate}

\author{P. P. Kizin$^1$, D. A. Zezyulin$^2$, and G. L. Alfimov$^1$}
\affiliation{$^1$Moscow Institute of Electronic Engineering,
Zelenograd, Moscow, 124498, Russia\\%
$^2$Centro de F\'isica Te\'orica e Computacional,
Faculdade de Ci\^encias da Universidade de Lisboa, Campo Grande, Edif\'icio C8, Lisboa  P-1749-016, Portugal\\ {dzezyulin@fc.ul.pt}
}

\begin{abstract}
The paper is devoted to numerical study of stability of nonlinear
localized modes (``gap solitons'') for the spatially one-dimensional Gross-Pitaevskii
equation (1D GPE) with periodic potential and repulsive interparticle
interactions. We use the Evans function approach combined with the
exterior algebra formulation in order to detect and describe weak
oscillatory instabilities. We show that the simplest (``fundamental'')
gap solitons in the first and in the second spectral gaps can undergo
oscillatory instabilities for certain values of the frequency parameter
(i.e., the chemical potential). The number of unstable eigenvalues and the
associated instability rates are described. Several stable and unstable
more complex (non-fundamental) gap solitons are also discussed.
The results obtained from the Evans function approach are
independently confirmed using the direct numerical integration
of the  GPE.
\end{abstract}

\maketitle

\section{Introduction}
\label{sec:intro}

The Gross--Pitaevskii equation (GPE),
\begin{eqnarray}\label{eq:gpe}
  i\Psi_t = -\Psi_{xx} + V(x)\Psi + \sigma|\Psi|^2\Psi,
\end{eqnarray}
describes the meanfield dynamics of a quasi-one-dimensional Bose-Einstein
condensate (BEC) confined in the potential $V(x)$ \cite{PS03}.
In Eq.~(\ref{eq:gpe}), $\Psi=\Psi( t, x)$  is the complex-valued   macroscopic wavefunction of the condensate.
The squared amplitude of the wavefunction $|\Psi(t, x)|^2$ describes
the local density of the BEC, while the gradient $(\arg\Psi( t, x))_x$
describes the velocity of atoms of condensate. The nonlinear term
$\sigma |\Psi|^2\Psi$ takes into account the interactions between the particles.
The case $\sigma=1$ corresponds to repulsive interparticle interactions,
while $\sigma =-1$ describes attractive interactions. Both these cases are
of physical relevance, but in what follows, we mainly focus on the
\emph{repulsive case}, $\sigma=1$. It is assumed that the
potential $V(x)$ is \emph{periodic}, which
corresponds to the  optical confinement of the BEC \cite{BK04, MO06}.

An important class of solutions of the GPE (\ref{eq:gpe}) corresponds to the stationary modes which  can be  represented  in the form
$\Psi(t,x) = e^{-i\mu t}u(x)$, where $\mu$ is a real parameter
 having the meaning of the chemical potential of the BEC. Function $u(x)$ satisfies the conditions of the spatial localization
\begin{eqnarray}\label{eq:bound_cond}
  \lim_{x\to\pm\infty} u(x)=0.
\end{eqnarray}
Without loss of generality one can assume that   $u(x)$ is   real-valued \cite{AKS02}. Then $u(x)$ can be found from the   stationary GPE
\begin{eqnarray}\label{eq:main}
  u_{xx} + (\mu - V(x))u - \sigma u^3=0.
\end{eqnarray}
If the potential $V(x)$ is periodic, the nonlinear modes satisfying
(\ref{eq:bound_cond})--(\ref{eq:main}) are called \emph{gap solitons}
\cite{BK04,  MO06, EK2003, HOM02, Kiv2003, PSK04},  since values of
$\mu$ corresponding to these solutions lie in the
spectral gaps of the linear Schr\"odinger equation \cite{BS91}
\begin{eqnarray}\label{eq:linear}
  u_{xx} + (\mu - V(x))u = 0.
\end{eqnarray}

\begin{figure}
  \includegraphics[width=0.46\textwidth]{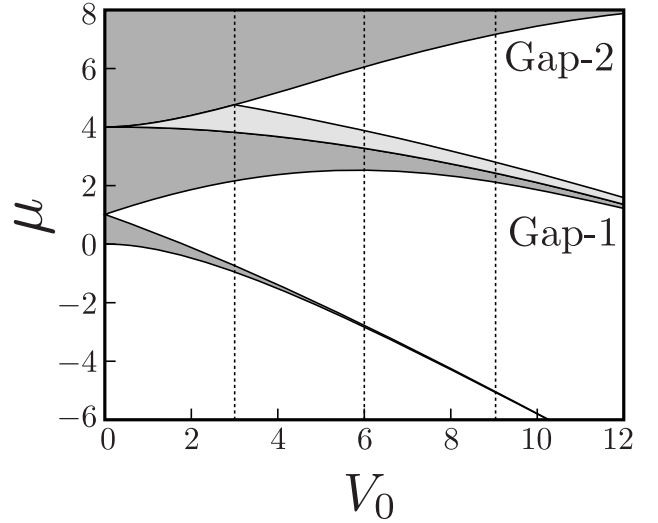}
  \caption{Band-gap structure of Eq.~(\ref{eq:linear}) with potential
    (\ref{eq:potential}). Dark gray regions represent bands of the
    spectrum. Light gray domain occupies the part of the second gap
    where the fundamental and multi-hump solitons do not exist (see
    Fig.~\ref{fig:gamma2}). Vertical dotted lines correspond to the
    potentials $V_0=3,6,9$   considered  in the present study.}
  \label{fig:bandgaps}
\end{figure}

The simplest class of the gap solitons are
\textit{fundamental gap solitons} (FGSs) \cite{EK2003,  Kiv2003,Zhang09_1,Zhang09_2, MM2006}. Under the repulsive nonlinearity, in the first gap there exists one family of FGSs  which   are characterized by the
presence of  a single dominating peak localized in one well of the potential $V(x)$. A variety of more complex (or \textit{higher-order}) solitons includes
truncated Bloch waves \cite{TrunkW09, Kiv2006} (which consist of
several in-phase peaks placed in a row),
various asymmetric
states, and  complex bound states of two (or more) well-separated waves
\cite{AHY12}, etc.   In spite of their rich
diversity, under certain (not very restrictive) conditions \textit{all} possible gap
solitons in a repulsive BEC
can be viewed as complexes of FGSs and classified using an alphabet consisting of a few symbols \cite{AA13}. Specifically, if the lattice is deep enough, then all the gap solitons in the first gap can be put into a one-to-one correspondence with the set of bi-infinite sequences of symbols from  a three-symbol alphabet. In simple terms, these symbols denote the presence or  the absence of the FGS (taken with plus or minus sign) in a potential well situated on the period of the potential $V(x)$.  For instance,
the truncated Bloch waves \cite{TrunkW09, Kiv2006} consisting of
several in-phase peaks placed in a row can be viewed as  complexes of single-hump FGSs. For classification of the gap solitons in the second spectral gap,
an alphabet of five symbols is necessary, and so on.

An important property of a gap soliton is its stability, since only dynamically
stable modes are likely to be experimentally feasible. The literature about
stability of gap solitons is rather abundant  \cite{HOM02,Kiv2003, PSK04,
Pelinovsky11, Zhang09_2, MM2006,TrunkW09, Yang10,  HAY11}. The most relevant
for our study outcomes for the case of repulsive interactions ($\sigma=1$)  can be summarized as follows. Significant part of the studies
concluded that  the single-hump FGSs are stable in the first gap  \cite{HOM02,Kiv2003,Zhang09_1,Zhang09_2,MM2006} and in the second gap
\cite{Zhang09_1,Zhang09_2,MM2006}.  Regarding the higher-order
states consisting of two or three in-phase peaks, they have been reported
unstable near the upper band edge in \cite{Kiv2003} in the first gap.
However, these states have been found to be stable both in the first
\cite{Zhang09_2,TrunkW09} and in the second \cite{Zhang09_2}
gap if the lattice depth is large enough.

\textcolor{black}{The  results listed above} have been obtained on the basis of
numerical studies of stability. In the meanwhile, it is recognized that the
numerical analysis of stability of the gap solitons is quite  a delicate
problem. A standard approach to the stability relies on the linear (or
spectral) stability technique which reduces the stability question to a
study of the spectrum of a certain linear operator. Depending on the
character of unstable eigenvalues, the instability typically manifests
itself either as a purely \emph{exponential instability} (when the
unstable eigenvalues   have zero imaginary parts) or as an
\emph{oscillatory instability} (OI) (when the unstable eigenvalues
are complex with nonzero imaginary parts). While the instabilities
of the former type are relatively simple to detect \cite{PSK04,Yang10},
the accurate  tracing of OIs is much more  challenging  \cite{Kiv2003,
PSK04}. As a result, the information about OIs of gap solitons in
optical lattices is rather scarce.
The absence of  information
on OIs for the simplest one-hump gap solitons in GPE is especially remarkable in view of well-known  OIs of Bragg gap solitons in nonlinear Dirac equations
\cite{BarPelZem98,BarZem00, Derk05}. The
latter system can be deduced from the GPE with a shallow periodic potential
using an asymptotic multiple-scale expansion \cite{Pelinovsky11,SM04}, and
therefore the results about OIs of the solitons for Dirac system seem  to be not consistent
with the stability results for the single-hump FGS mentioned above.

The numerical difficulties arising in the analysis of the OIs of the gap
solitons are related to  several issues. First, the rates of OIs are typically
quite small
\cite{PSK04,Yang10}. Another difficulty results from  poor localization
of the gap soliton and (or) of the eigenfunction associated with an unstable
eigenvalue.
This situation typically takes place when the chemical potential $\mu$ is close to the  gap edge. It requires unpractically wide computational windows or
a particularly accurate treatment of the boundary conditions. Some of
these difficulties can be overcome using the Evans function approach
which was employed in \cite{PSK04} to trace OIs of gap solitons in
the attractive condensates. It was further demonstrated in \cite{Blank}
that the numerically accurate evaluation of the Evans function requires a careful
treatment of the stiffness issue which arises for some values of the
complex argument of the Evans function. The stiffness problem can be
fixed if one redefined the Evans function using the exterior algebra
formalism \cite{Derk05,Blank}. This idea  has been developed into
a robust  numerical technique which was demonstrated to provide
reliable results even for relatively weak instabilities of gap solitons
\cite{Derk05,Blank}.

In the present paper, we use the Evans function approach combined
with the exterior algebra formulation in order to reveal and describe
weak OIs of FGS and higher-order gap solitons in the repulsive
BEC.  We focus on the first and second spectral gaps.
In each gap, we consider the single-hump FGS and
two higher-order solitons bearing two or three in-phase humps.   The
main outcomes of our numerical study can be outlined as follows.

\begin{itemize}
  \item[1.] In the first gap, all the considered solitons (including the
   single-hump FGS) are stable far from the upper band edge, but undergo
    OIs in the region near the upper band edge.
    The width of this instability
    region is quite significant: it occupies about 15\%-20\% of the
    width of the first gap.

  \item[2.] {In the second gap, all the considered solitons (including the  FGSs) are, in general, unstable due to 
  	OIs.  However, in a sufficiently deep potential, there exist intervals of $\mu$ where FGS are stable}.
\end{itemize}

To the best of our knowledge, our results constitute the first explicit
demonstration and detailed description of OIs for FGSs in the case of repulsive
interactions $\sigma=1$ in the
first and in the second gap. On the other hand, our results advance the
current understanding of the higher-order modes \cite{TrunkW09},
since we show that they undergo OIs even in a deep potential.

In order to confirm the linear stability results, we have also performed a
series of direct simulations of temporal behaviour of the solitons in  the GPE
(\ref{eq:gpe}). The results of these studies agree with the conclusions obtained from
the linear stability analysis and display the slow decay of unstable gap solitons and
the persistent evolution of the stable ones.

The rest of the paper is organized as follows.  In Sec.~\ref{sec:families} we briefly describe the families of gap solitons whose stability is the main subject of the present study. In Sec.~\ref{sec:stability} we present our main results on linear and nonlinear stability of the gap solitons. Section~\ref{sec:conclusion} concludes the paper.

\begin{figure*}
  \includegraphics[width=1\textwidth]{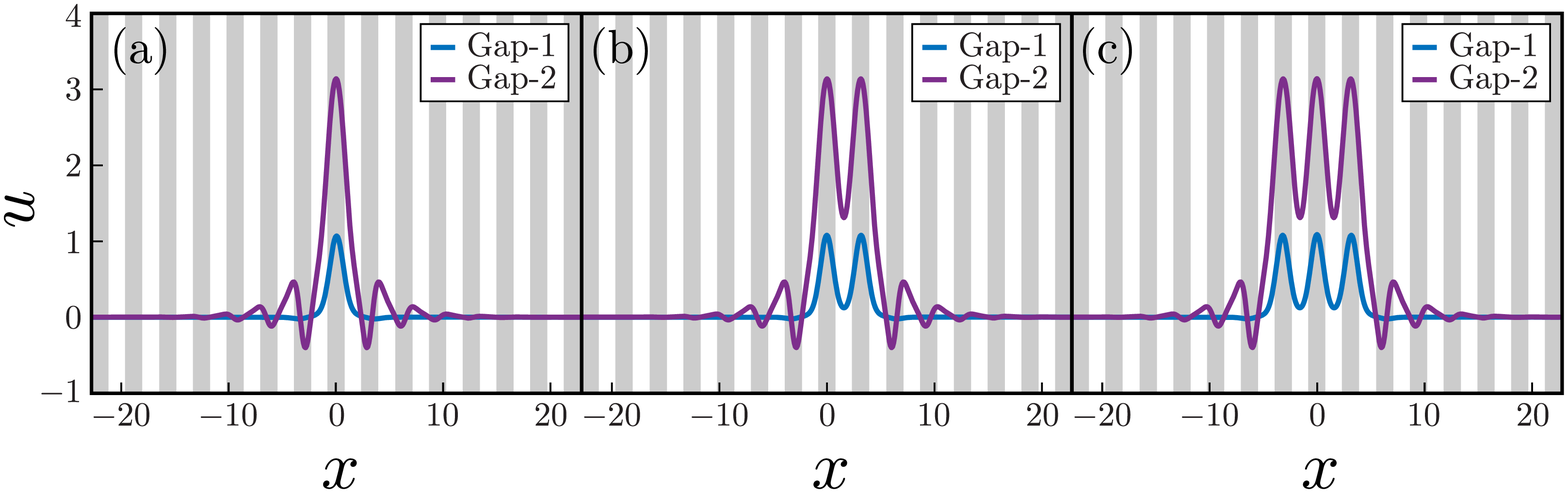}
  \caption{Examples of gap solitons in the first and the second gaps
    for the lattice with depth $V_0=6$. (a), (b) and (c) correspond to
    the one-hump FGS, two-hump and three-hump solitons,
    respectively. The chemical potential is  $\mu=-2$ and $\mu=5$ for solutions from the first and the  second gaps, respectively.
  }
  \label{fig:solutions}
\end{figure*}

\section{Families of gap solitons}
\label{sec:families}

In our study, we use a prototypical example of the periodic potential
in the form
\begin{eqnarray}\label{eq:potential}
  V(x) = - V_0\cos(2x),
\end{eqnarray}
where real $V_0>0$ characterizes the depth of the lattice.

The spectrum of the linear eigenvalue problem (\ref{eq:linear})
with the potential (\ref{eq:potential}) consists of one semi-infinite
gap and a countable set of finite gaps \cite{BS91}
(see Fig.~\ref{fig:bandgaps}). In our study, we consider the gap
solitons in the first and in the second gaps (no gap solitons exist
in the semi-infinite gap in repulsive BECs).
 In each of the gaps,
we consider three families of nonlinear modes: the fundamental
gap soliton with the single dominating peak at $x=0$ (the single-hump FGS),
and two higher-order solitons, with two and three dominating peaks. In
the terminology of \cite{TrunkW09}, these higher-order
solitons correspond to the truncated Bloch waves, and in terms of the
coding approach \cite{AA13} they are the bound states of two or three
in-phase single-hump FGSs with codes $(++)$ and $(+++)$. \textcolor{black}{Equation (\ref{eq:main}) also supports families of out-of-phase multi-hump solitons (with the codes  $(+-)$,  $(+-+)$, etc. However, these solutions have been reported to suffer  relatively strong exponential instabilities \cite{Yang10} and hence are not considered in our study.}

Representative
spatial profiles of the considered solutions for two different
values of the chemical potential $\mu$ are shown in
Fig.~\ref{fig:solutions}. Changing value of $\mu$ inside the
gap, it is possible to construct numerically continuous families
of gap solitons \cite{Kiv2003,  PSK04, Zhang09_2,
MM2006,TrunkW09,Yang10}. These families can be visualized in the
plane $N$ vs. $\mu$, where the squared $L^2$-norm
$N=\int_{-\infty}^\infty |u|^2 dx$ has the physical
meaning of  the number of particles in the condensate. The
dependencies $N$ vs. $\mu$ for $V_0=6$ are presented in
Fig.~\ref{fig:gamma}(a). In the first gap, one of the families
(corresponding to the single-hump FGS) bifurcates from the
edge of the spectral band. The two other curves do not bifurcate
from the band edge, \textcolor{black}{but appear as a result of  bifurcations which take place at a small but finite distance from the gap edge \cite{Yang10,AHY12}}. All the three curves can be continued to the upper edge of the first   gap.
\textcolor{black}{Inside the band, the  solitons do not exist.
However, fundamental and multi-hump solitons can be found again when  $\mu$ lies in the second gap.  The considered families do not bifurcate from the   edge of the second gap, but  exist only as $\mu$ exceeds a certain finite threshold  \cite{Zhang09_2} (the values of these thresholds for the three considered families turn out to be very close,  see the light  gray shading in  Fig.~\ref{fig:gamma} which shows the interval  in  the second gap where the families do not exist.)}
The shapes of the  solitons in the second gap are similar to their counterparts in the first gap, see Fig.~\ref{fig:solutions}.

\textcolor{black}{In the second gap, one can also find two families of gap solitons bifurcating from the gap edge (not shown in Fig.~\ref{fig:gamma}) \cite{PSK04,HAY11}. One of them is  exponentially unstable \cite{PSK04}, and another one (called  {\it subfundamental soliton} in \cite{MM2006}) is   unstable as the number of particles exceeds a certain threshold  \cite{Zhang09_2}. Since  instabilities of these solutions are well-known, in what follows we do not incorporate them in our study and focus on the simplest  FGS and solitons consisting of  two and three in-phase FGSs whose instabilities have not been seen before.
}

\section{Stability of gap solitons}
\label{sec:stability}

\subsection{Statement of the problem and the numerical method}

Following to the standard procedure of the linear stability analysis,
we consider a perturbed solution
\begin{eqnarray}\label{eq:perturb}
  {\Psi}(x,t)=({u(x)+e^{\lambda t}[a(x)+ib(x)]})
  \mathrm{e}^{-i\mu t},
\end{eqnarray}
where $u(x)$ describes a profile of a gap soliton, while $a$ and $b$,
$|a|, |b|\ll 1$, describe real and imaginary parts of a small-amplitude
perturbation. After substitution of (\ref{eq:perturb}) into the
GPE (\ref{eq:gpe}) and neglecting higher-order in $a$ and $b$ terms, one arrives at the following linear eigenvalue problem
\begin{eqnarray}\label{eq:LinStab}
  \lambda
  \begin{pmatrix}
    a \\ b
  \end{pmatrix} =L
  \begin{pmatrix}
    a \\ b
  \end{pmatrix}, \quad L =
  \begin{pmatrix}
    0 & -\mathcal{L}^- \\
    \mathcal{L}^+ & 0
  \end{pmatrix},
\end{eqnarray}
where linear operators $\mathcal{L}^\pm$ are defined as
\begin{eqnarray*}
  \mathcal{L}^{-} = \partial_x^2 + P - u^2, \quad
  \mathcal{L}^{+} = \partial_x^2 + P - 3u^2,
\end{eqnarray*}
and $P=\mu+V_0\cos 2x$. If the spectrum of $L$ is purely
imaginary, then the amplitudes of perturbations $a(x)$ and $b(x)$
do not grow, and the soliton is said to be linearly stable. On the other
hand, if at least one eigenvalue $\lambda$ has nonzero real part, then
the soliton is unstable. The largest real part of the eigenvalues characterizes the
instability growth rate.

\textcolor{black}{
The eigenvalue problem (\ref{eq:LinStab}) has a double eigenvalue
$\lambda=0$. The corresponding  eigenvector $v_1$ and the generalized eigenvector $v_2$ read
\begin{eqnarray}
\label{eq:ker}
v_1 =  \begin{pmatrix}
    0 \\ u
  \end{pmatrix},  \quad v_2 =  \begin{pmatrix}
  \partial_\mu u \\ 0
  \end{pmatrix}.
\end{eqnarray}
Then   $Lv_1 = 0$ and  $L^2v_2=0$. In (\ref{eq:ker}),
 $\partial_\mu u =\partial u(x;\mu)/\partial \mu$ is the partial derivative of the solution with respect to the chemical potential $\mu$  (provided that this derivative exists).}
It is also easy to see that if  $\lambda$ belongs to the spectrum of
the operator $L$, then $-\lambda$ and $\pm
\lambda^*$ also belong to the spectrum. Thus there are
 two typical scenarios of instability: (i) the
instability may take place due to the presence of a pair of real
eigenvalues $\lambda$ and $-\lambda$ and (ii) it may be caused by a
\textit{quartet} of unstable eigenvalues $(\lambda, \lambda^*,
-\lambda, -\lambda^*)$, where real and imaginary parts of
$\lambda$ are nonzero. The instabilities of the type (ii) are
known  as \emph{oscillatory instabilities} \cite{PSK04}.

In spite of recent advances in the rigorous theory for analysis of the
eigenvalue problem (\ref{eq:LinStab}) \cite{Pelinovsky11, KP13},
in the general case the description of its spectrum can be pursued
only numerically. Numerical solution of the linear stability problem
(\ref{eq:LinStab}) is recognized to be a sufficiently challenging
problem \cite{Kiv2003,PSK04}. One of well-elaborated numerical approaches to this task is based on the   the Evans function \cite{PSK04,KP13,Derk05,Blank}
which is especially useful if the eigenfunctions associated with unstable
eigenvalues of (\ref{eq:LinStab}) are poorly localized.  The Evans
function $f(\lambda)$ is an analytic function defined in the complex
domain except the points of the essential spectrum of the linear
stability operator $L$. The set of zeros of the Evans function
coincides with the set of the isolated eigenvalues of the linear stability
operator. Moreover, the order of the zero of the Evans function is
equal to the algebraic multiplicity of the corresponding isolated
eigenvalue.

According to the standard definition \cite{KP13}, the Evans function is a $\lambda$-dependent determinant whose  columns  are the 
vectors of stable and unstable manifolds for the linear stability problem
(\ref{eq:LinStab}).  In this case the problem (\ref{eq:LinStab})
is treated as a system of ODEs depending on the
complex parameter $\lambda$, and these vectors should be computed
solving the Cauchy problem for (\ref{eq:LinStab}).
However, it was demonstrated  \cite{Blank}  that for
certain values of the complex argument $\lambda$ the system
(\ref{eq:LinStab}) is stiff and the direct evaluation of the Evans function
cannot be fulfilled with  the sufficient accuracy. The stiffness issue can be overcome
by redefinition of the Evans function using the exterior algebra
formalism \cite{Derk05,Blank}. Redefined in this way, the Evans function was used
to study the stability of the surface gap solitons \cite{Blank}. In the
present study, we use  the exterior algebra formulation of the Evans function
to describe oscillatory instabilities of the fundamental and higher order
gap solitons in the repulsive BEC. While the detailed explanation of the numerical approach
can be found in \cite{Blank}, for the sake of self-containment of our
work  we  describe briefly the  main ingredients of the method  in the Appendix~A. Appendix~B addresses technical details of implementation of the method. Additionally, in   Appendix~B we briefly compare   the stability results obtained with the  Evans function approach and with  the Fourier collocation method which is another well-elaborated tool for computing the instabilities of gap solitons  \cite{Yang10}. As follows from our results, the accuracy of the Fourier collocation method may be not sufficient to compute  weak OIs. This confirms that the intricate Evans function technique is   essential for the accurate   tracing of  OIs of the FGSs.

\subsection{Linear stability results}

\begin{figure*}
  \includegraphics[width=0.85\textwidth]{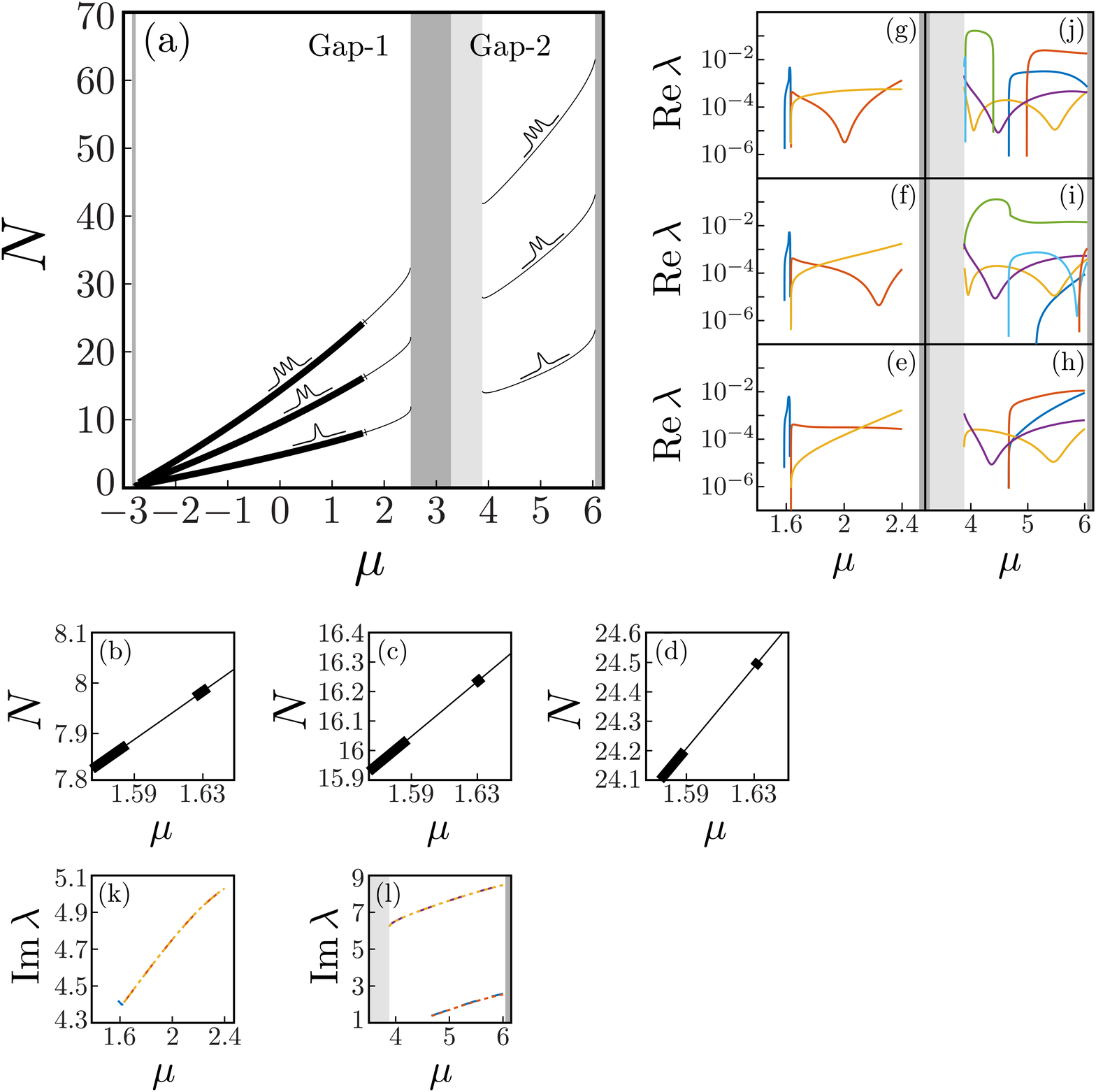}
  \caption{(a): Families of the single-hump FGS, two-hump and three-hump  gap solitons in the first and second spectral gaps for $V_0=6$ visualized as
    dependencies $N(\mu)$. The shapes of the solitons are shown schematically next to the curves. Bold fragments of the curves correspond
    to linearly stable solutions, while the thin fragments represent the
    unstable solutions. Dark gray domains are the spectral bands. Light gray shading  occupies  the part of the second gap where the considered families do not exist.
    Panels (b)-(d): magnification of  the narrow  regions of instability which exist in the families of single-hump FGS (b) and  multi-hump (c)-(d) solitons in the first gap. Panels (e)-(j):
    Reals parts of unstable eigenvalues. Panels (e) and (h)
    correspond to single-hump FGSs in the first  and in the second gaps, respectively;
    (f) and (i) correspond to the two-hump solitons; (g) and (j)
    correspond to the three-hump solitons. Panels (k) and (l): Imaginary
    parts of unstable eigenvalues whose real parts are shown in (e) and (h).   Notice that some of unstable eigenvalues illustrated in panels (e,k) and (h,l) have different real parts but  virtually equal imaginary parts which are not distinguishable on the scale of the panels; as a result,  the number of the curves visible in panels (l) and (k) is less than that in panels (e) and (h), respectively.}
  \label{fig:gamma}
\end{figure*}

\begin{figure*}
  \includegraphics[width=0.85\textwidth]{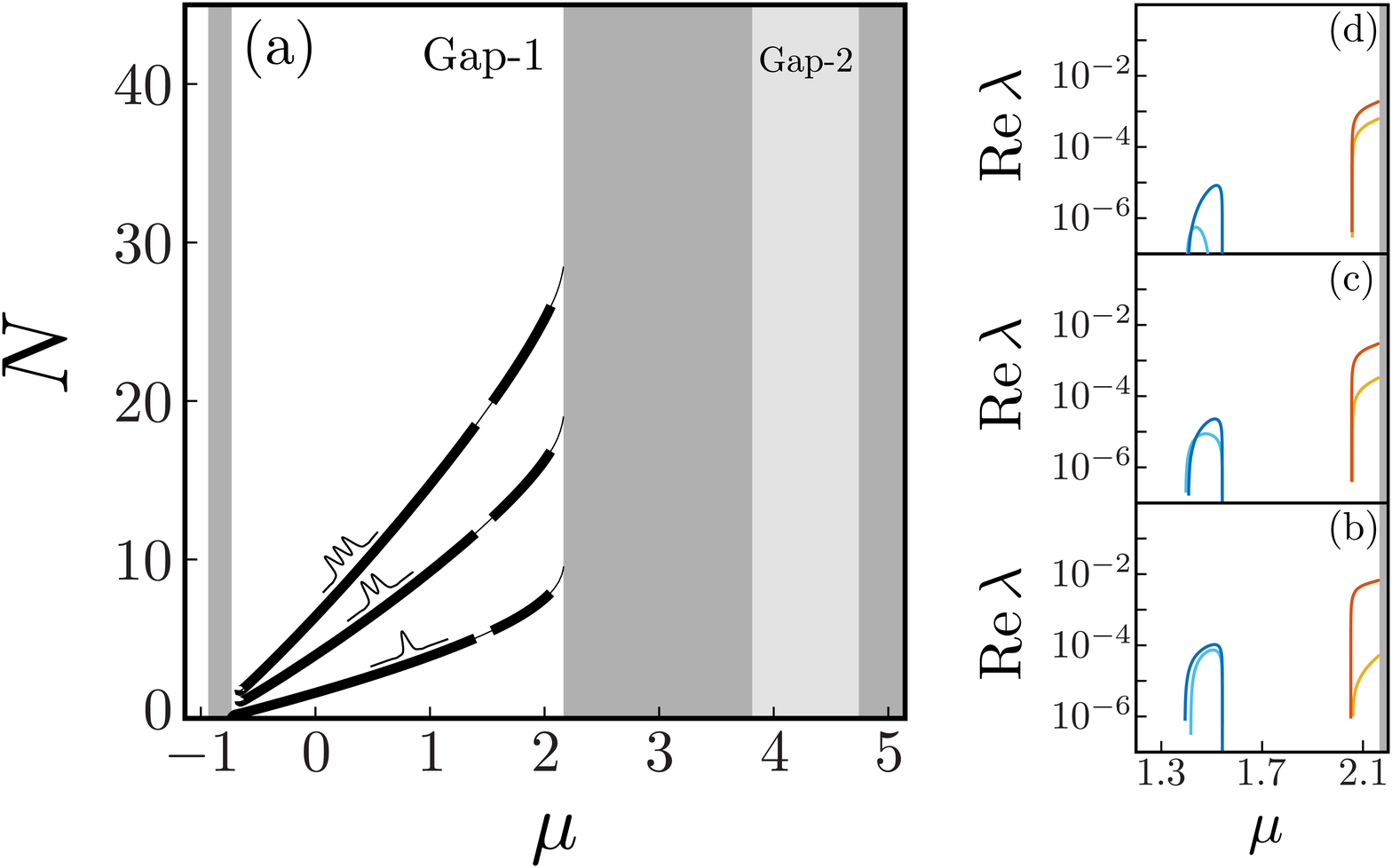}
  \caption{\textcolor{black}{
    Panels (a)-(c): Families of   FGSs (a), two-hump solitons (b), and three-hump solitons (c)  in  the shallow potential $V_0=3$.
 Dark gray domains are the spectral bands. The  families do not exist in the second gap, as indicated by the light gray shading which occupies the entire second  gap.
    Panels (b)-(d): Real parts of unstable eigenvalues   FGSs (b), two-hump (c), and three-hump (d)   solitons in the first gap.}}
  \label{fig:gamma2}
\end{figure*}

\begin{figure*}
	\includegraphics[width=0.85\textwidth]{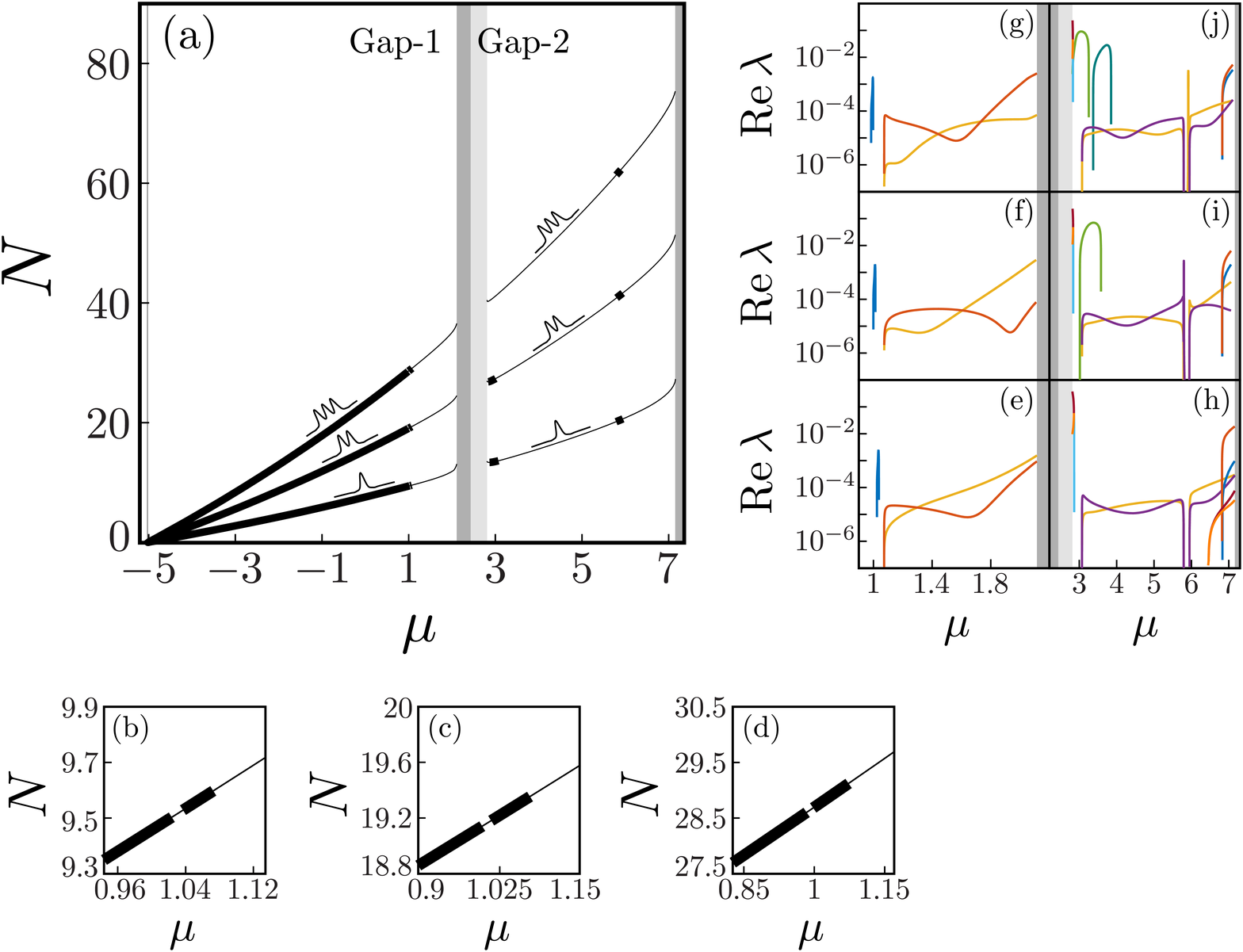}
	\caption{\textcolor{black}{
			Panels (a)-(c): Families of   FGSs, two-hump solitons, and three-hump solitons   in  the deep potential $V_0=9$.
			Dark gray domains are the spectral bands.  Light gray shading  shows the domain of the second gap where the considered families   do not exist. Panels (b)-(d): magnification of  the narrow  regions of instability which exist in the families of single-hump FGS (b) and  multi-hump (c)-(d) solitons in the first gap. Panels (e)-(j):
			Reals parts of unstable eigenvalues. Panels (e) and (h)
			correspond to single-hump FGSs in the first  and in the second gaps, respectively;
			(f) and (i) correspond to the two-hump solitons; (g) and (j)
			correspond to the three-hump solitons.}}
	\label{fig:gamma3}
\end{figure*}

Using the Evans function approach, we have examined in details  stability of   fundamental (single-hump) and multi-hump  solitons in the first and the second spectral gaps of a lattice (\ref{eq:potential})  with the fixed   depth  $V_0=6$. The results of this study are presented in Fig.~\ref{fig:gamma}. Then, in order to check that the obtained results are generic and to understand the effect of the depth of the lattice on the found instabilities, we have repeated  the computations  in a more shallow ($V_0=3$) and a deeper ($V_0=9$) lattices, see  Fig.~\ref{fig:gamma2} and Fig.~\ref{fig:gamma3}. The chosen depths of the potential are shown with vertical dashed lined in Fig.~\ref{fig:bandgaps}.  Let us now  proceed to the detailed presentation
of the main outcomes of our work.


Figure~\ref{fig:gamma}  shows the   results of the first series of numerical
experiments. Let us start from the FGSs in the first gap. As shown in
Fig.~\ref{fig:gamma}(e), we have not found any {unstable eigenvalue} of FGS for $\mu<1.586$. Therefore, we conjecture that the FGSs are stable for the corresponding  values of the chemical potential. However, for larger $\mu$,
we have identified three quartets of  {unstable eigenvalues. Imaginary parts of the unstable eigenvalues are typically
nonzero, which confirms to the oscillatory character of the instabilities
[see Fig.~\ref{fig:gamma}(k)]}. The first quartet of unstable eigenvalues exists only
in a relatively narrow window $1.586\lesssim\mu \lesssim 1.625$ [Fig.~\ref{fig:gamma}(b)].
The other two coexist in a sufficiently wide instability window
which starts at  $\mu\approx 1.633$  at continues up to the
upper gap edge. In the interval $1.625 \lesssim \mu \lesssim1.633$ no instability
has been found [see Fig.~\ref{fig:gamma}(b)]. Looking at the instability rates associated with the found OIs [Fig.~\ref{fig:gamma}(e)], we
notice that the real parts of the found   eigenvalues $\textrm{Re}\lambda$
do not exceed $10^{-2}$ and typically lie between $10^{-5}$ and
$10^{-3}$.

Turning to multi-hump solitons in the first gap [Fig.~\ref{fig:gamma}(f,g)], we  observe that the stability
picture is remarkably similar  to that for the FGSs: no unstable eigenvalues is observed for sufficiently small $\mu$, but for large values of the chemical potential the solutions are  unstable   due to
three quartets of unstable eigenvalues which cause   two windows of instability. For the family of
two-hump solitons,  the onset of instability  is  detected  at $\mu \approx 1.628$. The instability persists in the
narrow interval $1.588\lesssim\mu\lesssim1.628$ [see Fig.~\ref{fig:gamma}(c)].  \textcolor{black}{No unstable eigenvalues is found  for  $1.628\lesssim\mu
\lesssim1.633$}, but the instability again takes place for all $\mu$
between $1.633$ and the upper gap edge. In a similar way, the
three-hump solitons have two windows of instability: the
narrow window is situated at $1.589\lesssim\mu\lesssim1.630$ [Fig.~\ref{fig:gamma}(d)],
and the wide interval starts  at  $\mu\approx 1.633$ and
continuous up to gap edge. The typical instability rates of
the multi-hump solitons are comparable with those  for  the
fundamental solitons and do not exceed $10^{-2}$.

Proceeding to the solitons in the second gap, we observe that all
three considered solutions are unstable in the entire domain of their
existence. The one-hump FGSs [Fig.~\ref{fig:gamma}(h)] are
unstable due to the presence of two (or more, depending on the
particular value of $\mu$)  {quartets of unstable eigenvalues}. Imaginary parts of these eigenvalues are presented in
Fig.~\ref{fig:gamma}(l). Again, the instability rate is below
$10^{-2}$. However, the instabilities become stronger for multi-hump
solutions in the second gap [Fig.~\ref{fig:gamma}(i, k)]:
for these solutions $|\textrm{Re}
\lambda|$  reaches $10^{-1}$. In general, multi-hump
solitons are unstable because of the  presence of {\emph{multiple} coexisting quartets of  unstable eigenvalues}.

In order to examine the role of the lattice  depth, we performed the second series of numerical runs extending {the study  onto    solitons in a more shallow ($V_0=3$) and  a  deeper ($V_0=9$) potentials.
Figure~\ref{fig:gamma2} presents the stability results for the shallow potential.}
We observe that the solitons in the first gap also feature two windows of instability, similar to their counterparts  with $V_0=6$. For the fundamental family the first window of instability is  situated at $1.394\lesssim \mu\lesssim 1.542$.
In this region, the spectrum includes two  {quartets of unstable eigenvalues}. %
The second window   of instability  starts at $\mu\approx 2.052$  and continues to
the upper gap edge. In this case, the instability is again associated with
two quartets of eigenvalues. For two-hump and three-hump solitons  the picture is qualitatively the same.
The considered  families of gap solitons   do  not exist in the second gap of the shallow
lattice, and {therefore the corresponding panels with the stability results are absent  Fig.~\ref{fig:gamma2}.}

Turning to the deep lattice, $V_0=9$ [see Fig.~\ref{fig:gamma3}],
we again observe two intervals of instability of solitons   in the first gap.
The instability rate   remains small and typically does not exceed
$10^{-3}$. More interestingly, we \textcolor{black}{can conjecture}  that solitons in the \textit{second gap}  are
stable if $\mu$ belongs to  certain narrow stability windows    [Fig.~\ref{fig:gamma3}(h--j)]. For example, for the FGSs in the second gap, no unstable eigenvalues has been found  near the lower gap
edge $\mu \lesssim  3.072$ and in the narrow window $5.793 \lesssim
\mu \lesssim 5.949$. Thus the increase of $V_0$  enhances the stability of FGSs in
the second gap.

\subsection{Solution of the time-dependent GPE}

\begin{figure}
	\includegraphics[width=0.8\columnwidth]{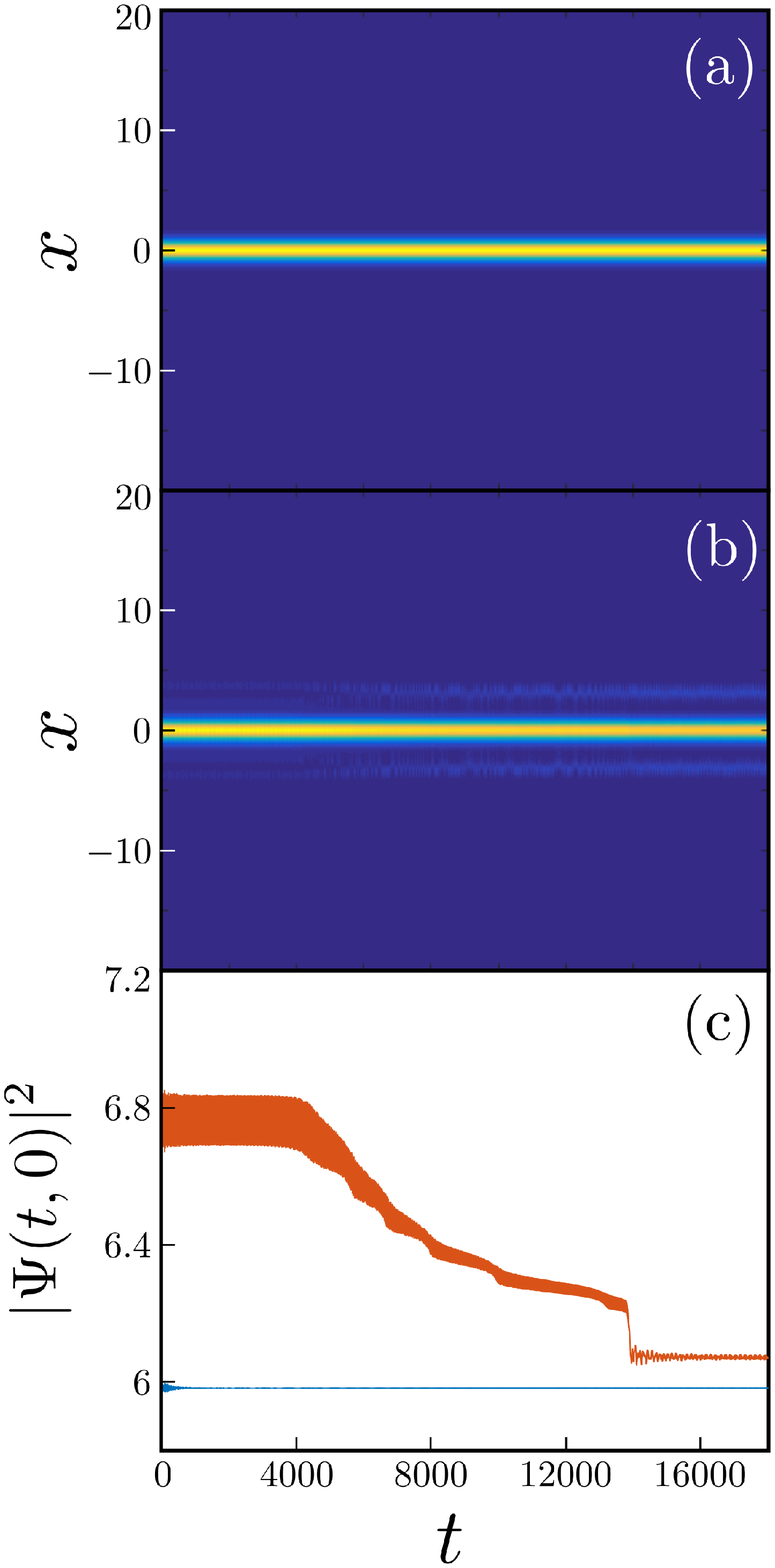}
	\caption{Evolutions of two one-hump FGSs  in the first gap, $V_0=6$: stable soliton with $\mu=1.5$ and unstable soliton with $\mu=2.2$.
		Panels (a) anb (b): Pseudo-color plots of the amplitude $|\Psi(t,x)|^2$
		of Eq.~(\ref{eq:gpe}) with $\mu=1.5$ (a) and $\mu=2.2$ (b). The oscillatory instability in panel (b) manifests itself in the developing small-amplitude oscillations which are especially well-visible at the soliton tails. Panel (c) shows the dependencies $|\Psi(t, x=0)|^2$ for stable (blue curve) and unstable (red curve)  solutions.
	}
	\label{fig:evol1}
\end{figure}

\begin{figure}
	\includegraphics[width=0.8\columnwidth]{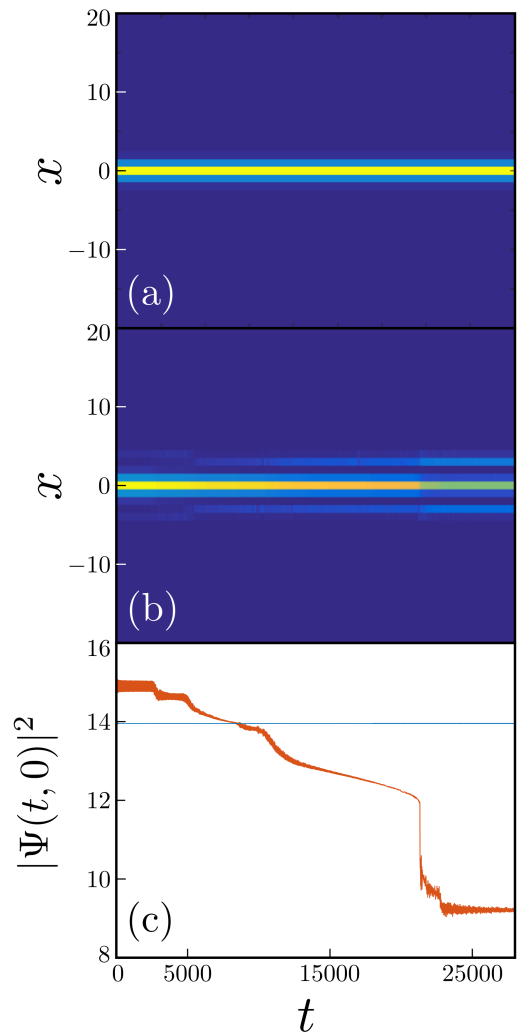}
	\caption{Evolutions of two one-hump FGSs in the second gap of the deep lattice, $V_0=9$: stable soliton with $\mu=5.85$ and an unstable soliton with $\mu=7$. Panels (a) and (b): Pseudo-color plot of stable and unstable evolutions, respectively. Panel (c) shows the dependencies $|\Psi(t, x=0)|^2$ for stable (blue curve) and unstable (red curve)  solutions.
	}
	\label{fig:evol2}
\end{figure}

In order to check the linear stability results obtained with the Evans
function method, we have also tested the nonlinear stability of the gap
solitons by means of  the  direct integration of the GPE (\ref{eq:gpe}) using  a semi-implicit finite difference scheme
\cite{TroPes2009}. The equation was spatially discretized in a
sufficiently large computational window $x\in [-20\pi, 20\pi]$
with the grid step equal to $\pi/256$. We have additionally
introduced absorbing (dissipative) boundary  layers by adding nonzero imaginary
part to the potential $U(x)$ for $x>50$ and $x<-50$. These
absorbing layers mimic a setup in which the perturbations
are removed from the system as they escape from the central region and reach the boundaries. In order to boost the development of eventual instabilities, the initial conditions have been taken in
the form of slightly perturbed gap solitons. Specifically, we have
used a $3\%$-multiplicative perturbation: $\psi \to 1.03 \psi$.
The temporal step was $\tau = 10^{-3}$.

In general, the results of the evolutional simulations are in the
complete agreement with the predictions of the linear stability
analysis. For stable solutions, the amplitude of the introduced
perturbation does not grow, and the solution persists undistorted
for indefinitely long time. For unstable solutions, the introduced
perturbation grows slowly, which eventually leads to the breakup of the solution. In order to exemplify this pattern, in Fig.~\ref{fig:evol1}(a,b)
we compare the
dynamics of a stable and an unstable FGSs from the first gap. The
drastic difference in their evolutions is especially well-visible in the
plot with  $|\Psi(t,x=0)|^2$, see Fig.~\ref{fig:evol1}(c). To illustrate the dynamics in the second gap, in Fig.~\ref{fig:evol2} we compare the behavior of a stable FGS which exists in the second gap if the potential is deep enough [specifically, we chose $V_0=9$ as in Fig.~\ref{fig:gamma2}(b)] with an unstable FGS with the same $V_0$ but different $\mu$.
Again, the striking  difference in their behaviors is much better pronounced on the plot which shows the density at $x=0$ [Fig.~\ref{fig:evol2}(c)]. \textcolor{black}{Looking at the long-term dynamics of   solutions   in Figs.~\ref{fig:evol1} and \ref{fig:evol2}, we observe that amplitude of   unstable solutions  decreases, and after a long transient period the shape of solutions features small-amplitude oscillations around some localized states. These oscillations persist for indefinitely long time.}

The results for dynamics of multi-hump solitons also agree with the predictions of the linear stability analysis.

\section{Conclusion}
\label{sec:conclusion}

To conclude, we have performed a systematic  study of instabilities
of gap solitons in a repulsive Bose-Einstein condensate trapped in a
periodic potential. While this topic has been addressed in several
previous studies, the complete picture of stability of gap solitons have not been completely understood yet.  One of the reasons for this is related to the fact that the accurate study of instabilities of gap solitons can be a challenging problem which requires a sufficiently advanced numerical tool. In our study we used the Evans function approach combined with the external algebra formalism. The main advance made by our work is related to
the explicit presentation of oscillatory instabilities of the
\emph{fundamental solitons}. While the possibility of existence
of these instabilities has been hypothetically suggested, we present
the first, to the best our knowledge,  explicit demonstration of
unstable eigenvalues, describe the regions of stability and instability
and discuss the typical instability rates. Next, we have demonstrated that the found oscillatory instabilities are rather generic and can be also found in a more shallow and a deeper optical lattices. On the other hand, it was in particular found that the increase of the lattice depth can enhance the stability of FGSs
in the second gap: the entire branch of solutions is unstable for the depth equal to
$V_0=6$, but displays windows of stability for $V_0=9$.


Besides the fundamental single-hump solitons, we have also
examined stability of two-hump and three-hump solutions and
demonstrated that such solutions from the first gap suffer
oscillatory instabilities even in a deep potential.
Finally, we have performed a series of numerical runs with the direct integration of the
Gross-Pitaevskii equation and observed how  the predicted oscillatory instabilities manifest themselves during the temporal evolution.

\textcolor{black}{In our opinion, the results of this work may be regarded as a  first step in the study of  the hypothetical  relation between the code of a  gap soliton  (in sense of the paper \cite{AA13}) and the spectrum of the corresponding stability problem. For the solitons from the first bandgap,  the possibility of this relation  has been discussed in \cite{Yang10}, focusing mainly on the exponential instabilities of solitons coded by  heuristically constructed sequences  of ``$+$'' and ``$-$''   symbols. Another potential subject  for  further studies  is the analysis of bifurcations of gap solitons, in terms of their codes, which can be also  strongly related to their stability properties. Here we would mention the recent result of \cite{FS14} about the connection between the modes of Discrete Nonlinear Schrodinger equation and gap solitons of the GPE. This connection allows to identify the codes of solitons that ``annihilate''  each other, see \cite{ABK04} for the table of such bifurcations.}

\section*{Acknowledgments}

The work of DAZ was supported by the FCT (Portugal) through the grant No. UID/FIS/00618/2013.

\appendix

\section{Definition of the Evans function}

Following to \cite{Blank}, we rewrite the linear stability
eigenvalue problem  (\ref{eq:LinStab}) as follows
\begin{equation}\label{eq:LinStabU}
  \lambda
  \begin{pmatrix}
    p \\
    q
  \end{pmatrix} =i
  \begin{pmatrix}
    \partial_x^2+P -2u^2  &  -u^2 \\
    u^2  &  -\partial_x^2-P + 2u^2
  \end{pmatrix}
  \begin{pmatrix}
    p \\
    q
  \end{pmatrix},
\end{equation}
where $p=(a+ib)/2$ and $q=(a-ib)/2$. Further, Eq.~(\ref{eq:LinStabU}) can be rewritten as an
ODE system, where $\lambda$ plays the role of a parameter:

  \begin{eqnarray}\label{eq:c4}
    {\bf u}_x=A{\bf u}, \quad
    {\bf u}= \textrm{col}(p,\, p_x, \, q, \, q_x),
  \end{eqnarray}
\begin{equation*}
A=
    \begin{pmatrix}
      0  &  1  &  0  &  0  \\
      2u^2-(P+i\lambda)  &  0  &  u^2  &  0  \\
      0  &  0  &  0  &  1  \\
      u^2  &  0  &  2u^2-(P-i\lambda)  &  0
    \end{pmatrix}.
\end{equation*}

\noindent Due to the localization $\lim_{x\to\pm\infty}{u(x)}
=0$, in the limit $x\to \pm \infty$ Eq.~(\ref{eq:c4}) decouples
into a pair of  Mathieu equations

\begin{align}
	&{\bf v}_x=B{\bf v}, \quad
	B=
	\begin{pmatrix}
		0 				& 1 		\\
		-(P+i\lambda) 	& 0
	\end{pmatrix},	\label{eq:Mathie1} \\
	&{\bf w}_x=C{\bf w}, \quad
	C=
	\begin{pmatrix}
		0 				& 1 		\\
		-(P-i\lambda) 	& 0
	\end{pmatrix}.	\label{eq:Mathie2}
\end{align}

Equation~(\ref{eq:Mathie1}) has the stable manifold ${\bf v}^+=
\left(v^+,v_x^+\right)^{\rm T}$  for $x\to \infty$ and the
unstable manifold ${\bf v}^-=\left(v^-,v_x^-\right)^{\rm T}$.
The stable and unstable manifolds can be found numerically using
the monodromy matrix associated with Eq.~(\ref{eq:Mathie1}).
In a similar way, one can compute the monodromy matrix associated
with Eq.~(\ref{eq:Mathie2}) and  identify its stable ${\bf w}^+=
\left(w^+,w_x^+\right)^{\rm T}$   and unstable ${\bf w}^-=
\left(w^-,w_x^-\right)^{\rm T}$ manifolds.
Using vectors  $\bf \{v^\pm\}$ and $\bf \{w^\pm\}$, we can find an unstable manifold $E^u$ of Eq.~(\ref{eq:c4})
and its stable  manifold $E^s$:
\begin{align}\label{eq:unstab}
  E^u (x;\lambda) &\sim
  \begin{pmatrix}
    v^+  &  0	 \\
    v_x^+  &  0  \\
    0  &  w^+  \\
    0  &  w_x^+	
  \end{pmatrix}
  \quad {\rm as}\quad x \to -\infty, \\
  \label{eq:stab}
  E^s (x;\lambda) &\sim
  \begin{pmatrix}
    v^-  &  0  \\
    v_x^-  &  0  \\
    0  &  w^-  \\
    0  &  w_x^-
  \end{pmatrix}
  \quad {\rm as}\quad x \to +\infty.
\end{align}
Using (\ref{eq:unstab})--(\ref{eq:stab}) the Evans function is
defined as the determinant of a $4\times 4$ matrix \cite{Blank}
\begin{equation}
	\label{eq:EvansC4}
	f(\lambda) = \frac{-1}{4\lambda+1} {\rm det}
	\Big[E^u (0; \lambda), E^s (0; \lambda)\Big].
\end{equation}

The definition (\ref{eq:EvansC4}) is, in principle, suitable for
numerical evaluation of the Evans function for different values of
the complex argument $\lambda$. To this end, one can start from a
sufficiently large $x=\pm L$, $L\gg 0$, solve the ODE system
from $-L$ to $0$ and from $L$ to $0$ in order to obtain the
necessary values at $x=0$, and compute the necessary determinant. However,
this procedure suffers from a stiffness problem which manifests itself
when the growth/decay rates of the two vectors in the manifold of
interest (stable or unstable) are largely different. This issue can be
overcome by a reformulation of (\ref{eq:c4})  in the exterior algebra.
Following to \cite{Blank}, we introduce a $6\times6$ matrix function

\begin{widetext}
  \begin{equation}
    A^{(2)}(x;\lambda)=
    \begin{pmatrix}
      0  &  u^2  &  0  &  0  &  0  &  0  \\
      0  &  0  &  1  &  1  &  0  &  0  \\
      0  &  2u^2-(P-i\lambda)  &  0  &  0  &  1  &  0  \\
      0  &  2u^2-(P+i\lambda)  &  0  &  0  &  1  &  0  \\
      -u^2  &  0  &  2u^2-(P+i\lambda)  &  2u^2-(P-i\lambda)  &  0  &  u^2  \\
      0  &  -u^2  &  0  &  0  &  0  &  0
    \end{pmatrix},
  \end{equation}
  \end{widetext}

\noindent and instead of solving system  (\ref{eq:c4}), we
introduce a new  ODE system
\begin{eqnarray}\label{eq:c6}
  {\bf U}_x=A^{(2)}{\bf U}.
\end{eqnarray}
We compute two solutions of (\ref{eq:c6}), ${\bf U}^u$ and
${\bf U}^s$, which are correspond to initial conditions
\begin{equation*}
	{\bf U}^u (-L;\lambda)=
		\begin{pmatrix}
			0 \\ v^+w^+ \\ v^+w_x^+ \\
			v_x^+w^+ \\ v_x^+w_x^+ \\ 0
		\end{pmatrix}, \quad
	{\bf U}^s (L;\lambda)=
		\begin{pmatrix}
			0 \\ v^-w^- \\ v^-w_x^- \\
			v_x^-w^- \\ v_x^-w_x^- \\ 0
		\end{pmatrix}.
\end{equation*}
Finally, the Evans function is redefined as
\begin{eqnarray}\label{eq:EvansC6}
  f(\lambda) = \frac{-1}{4\lambda+1}
  \Big\langle \overline{{\bf U}^u (0; \lambda)},
  \Sigma{\bf U}^s (0; \lambda)\Big\rangle,
\end{eqnarray}
where
\begin{eqnarray}
  \Sigma=
  \begin{pmatrix}
    0  &  0  &  0  &  0  &  0  &  1  \\
    0  &  0  &  0  &  0  &  -1  &  0  \\
    0  &  0  &  0  &  1  &  0  &  0  \\
    0  &  0  &  1  &  0  &  0  &  0  \\
    0  &  -1  &  0  &  0  &  0  &  0  \\
    1  &  0  &  0  &  0  &  0  &  0
  \end{pmatrix},
\end{eqnarray}
and  $\langle\cdot,\cdot\rangle$ stands for  the standard inner
product in $\mathbb{C}^6$ with the complex conjugation of the first argument.

\begin{center}
	\begin{table*}
		\begin{small}
			\begin{tabular}{|l||l|l|l|}
				\hline
				\backslashbox{$h$}{\\$L$}
				&  \makebox[3em]{$5\pi$}  &  \makebox[3em]{$10\pi$}
				&  \makebox[3em]{$15\pi$}  \\
				\hline
				&&&\\[-8pt]
				$\pi/64$  &  $-2.050526725+0.000827075i$  &
				$-2.081973594+0.000839757i$  &  $-2.113902729+0.000852636i$ \\
				$\pi/128$  &  $-0.183319425+0.000002354i$  &
				$-0.183359201+0.000002354i$  &  $-0.183399156+0.000002355i$\\
				$\pi/256$  &  $-0.013029289+0.000000005i$  &
				$-0.013029157+0.000000005i$  &  $-0.013029006+0.000000005i$\\
				$\pi/512$  &  $-0.000860190$  &  $-0.000860114$  &
				$-0.000860162$\\
				$\pi/1024$  &  $-0.000055153$  &  $-0.000055170$  &
				$-0.000055150$\\
				$\pi/2048$  &  $-0.000003552$  &  $-0.000003497$  &
				$-0.000003476$\\
				$\pi/4096$  &  $-0.000000280$  &  $-0.000000255$  &
				$-0.000000224$\\[2pt]
				\hline
			\end{tabular}																							
			\vspace{10pt}  \\
		\end{small}
		\caption{The numerically obtained value of the Evans function  $f(\lambda)$ at $\lambda=0$ for different $h$ and $L$; the computations correspond to the  FGS  from the first gap with
			$\mu=1.615$, $V_0=6$.}
		\label{tbl:acc}
	\end{table*}
\end{center}

\begin{center}
  \begin{table*}
  \begin{small}
    \begin{tabular}{|l||l|l|l|}
      \hline
      \backslashbox{$h$}{\\$L$}
      &  \makebox[3em]{$5\pi$}  &  \makebox[3em]{$10\pi$}
        &  \makebox[3em]{$15\pi$}  \\
      \hline
      &&&\\[-8pt]
      $\pi/128$  &  $0.005911869+4.398983883i$  &
        $0.005920080+4.398982678i$  &  $0.005920749+4.398982395i$\\
      $\pi/256$  &  $0.005920718+4.398988070i$  &
        $0.005922469+4.398990060i$  &  $0.005922665+4.398990200i$\\
      $\pi/512$  &  $0.005899840+4.398979139i$  &
        $0.005899988+4.398979328i$  &  $0.005900005+4.398979342i$\\
      $\pi/1024$  &  $0.005897874+4.398978297i$  &
        $0.005897885+4.398978310i$  &  $0.005897886+4.398978311i$\\
      $\pi/2048$  &  $0.005897731+4.398978235i$  &
        $0.005897732+4.398978236i$  &  $0.005897732+4.398978236i$\\
      $\pi/4096$  &  $0.005897722+4.398978231i$  &
        $0.005897722+4.398978231i$  &  $0.005897722+4.398978231i$\\[2pt]
      \hline
    \end{tabular}																							
    \vspace{10pt}  \\
  \end{small}
  \caption{Simple zero $\lambda$ of the Evans function $f(\lambda)$  obtained numerically with the Newton method for different $h$ and $L$.  The computations correspond to the same solution as in Table~\ref{tbl:acc}.}
  \label{tbl:acc2}
  \end{table*}
\end{center}

\section{Details of numerical implementation and accuracy issues}

The   evaluation of  the Evans function involves two
parameters of the numerical method: $h$ (step of the Runge-Kutta method) and $L$ (half-width of the
computational domain).
Besides, it requires a profile of the soliton $u(x)$.  Let us fix parameters of the
model $\mu=1.615$ and $V_0=6$ and perform the accuracy tests for the single-
hump FGS which exists with these parameters [see Fig.~\ref{fig:gamma}(a)].  In order to find the solution $u(x)$, we use the shooting method which
requires integration of the equation (\ref{eq:main}). The Runge-Kutta method with
the accuracy
$\mathcal{O}(h^4)$ was used.  In order to compute the Evans function, we solve
the ODE Eq.~(\ref{eq:c6}) using the Runge-Kutta method with the same accuracy.
In order to check the accuracy of
evaluation of the   Evans function, we use the fact that $\lambda=0$
is a double eigenvalue of  Eq.~(\ref{eq:LinStabU}), and therefore
\begin{equation}
  f(0)=f_\lambda(0)=0,  \quad
  f_{\lambda\lambda}(0)\ne 0.
\end{equation}
In Table~\ref{tbl:acc} we show the numerical value of the Evans function at $\lambda=0$. It  converges to the exact value $f(0)=0$ as $\mathcal{O}(h^4)$, which corresponds to the accuracy of the Runge-Kutta method.

In order to locate the zeros of the Evans function, we use the  Newton method. \textcolor{black}{The derivative of the Evans function with respect to $\lambda$ was approximated numerically using the first difference.}  As an example, in Table~\ref{tbl:acc2} we show the convergence to a simple zero computed for the same solution but with different $h$ and $L$.
For sufficiently small $h$ the  order of convergence is  close to $\mathcal{O}(h^4)$.

\textcolor{black}{While the Newton method is suitable for detecting and tracing of unstable eigenvalues, it, strictly speaking, does not ensure that \textit{all} unstable eigenvalues have been found; we therefore cannot give a ``computational proof'' of stability;  we can only conjecture on stability of a solution if no unstable eigenvalues has been found}.

Most   of the stability results presented above in Fig.~\ref{fig:gamma} and \ref{fig:gamma2} have been obtained with typical values  $h=\pi/256$,
$L=10\pi$ [for  computation of solitons and the  Evans function in the first spectral gap] and $L=20\pi$
[for  computation of solitons and   the Evans function in the second  gap. We have also selectively tested several particular unstable solitons by repeating the computation with smaller step $h$ and confirmed that the found instabilities are reproduced.

Finally,  we notice that the detection of OIs is indeed a complex problem, and an advanced numerical tool is necessary for its solution. To this end, let us compare the results of our analysis with the results obtained with  the Fourier collocation method  (FCM) \cite{Yang10} which is another common tool for study of spectral stability of gap solitons. The FCM is known to provide excellent results for detection of relatively strong and exponentials instabilities,  but as follows from our numerical experiments it may not be sufficiently robust to trace weak OIs.   To illustrate this,   let us consider  FGSs in the first gap with  $V_0=6$ fixing the parameters of the numerical method as $L=15\pi$, $h=\pi/4096\approx 10^{-3}$.  We consider three close values of the chemical potential $\mu$, namely,   $\mu=2.22, \mu=2.23$ and $\mu=2.25$. Using the  FCM (implemented with $300$ spectral harmonics using the computer codes from \cite{Yang10}) we have detected a pair of OIs ($\lambda_1=0.003670+4.934713i$ and $\lambda_2=0.005738+4.936358i$)   at $\mu=2.23$. However, the FCM  does  not indicate any instability for   $\mu=2.22$ and $\mu=2.25$. Moreover, if we try to check   the obtained results by repeating the computation in a smaller ($L=14\pi$) or in a larger ($L=20\pi$) computational window,    the FCM   does not find any unstable eigenvalue at all (for each of the three considered values of $\mu$). In a similar way, the results of the FCM are  sensitive to the choice of the spatial step $h$.  Hence, in this particular situation the results of the FCM cannot be considered as reliable.

On the other hand, the Evans function  indicates a pair of OIs for each of the chosen values of $\mu$. Namely, for $\mu=2.22$ the Evans function finds unstable eigenvalues $\lambda_1=0.000299+4.924631i$, $\lambda_2=0.000566+4.925482i$, for   $\mu=2.23$ it gives $\lambda_1=0.000298+4.931506i$, $\lambda_2=0.000601+4.932380i$, and at $\mu=2.25$ it gives $\lambda_1=0.000294+4.944917i$, $\lambda_2=0.000678+4.945838i$ [see  Fig.~\ref{fig:gamma}(e)]. Moreover,  these instabilities are reproduced with good accuracy if one repeats the numerical procedure with different $L$ and  $h$.  Thus the Evans function robustly traces the instability.


\end{document}